\documentclass[12pt,preprint]{aastex}
\usepackage{amsmath, amssymb}
\shorttitle{Soft Component and QPO of SAX J2103.5+4545}
\shortauthors{Inam et al.}
\begin{document}
\title{Discovery of Soft Spectral Component and Transient 22.7s Quasi Periodic 
Oscillations of SAX J2103.5+4545}
\author{S. \c{C}. \.{I}nam and A. Baykal}
\affil{Department of Physics, Middle East Technical University, Inonu
Bulvari, Balgat, Ankara 06531 Turkey}
\email{scinam@metu.edu.tr, altan@astroa.physics.metu.edu.tr}

\author{J. Swank}
\affil{NASA Goddard Space Flight Center, Greenbelt, MD 20771 USA}
\email{swank@xenos.gsfc.nasa.gov}
\and

\author{M.J.Stark}  
\affil{Department of Physics, Lafayette College, Easton, PA 18042 USA}
\email{starkm@lafayette.edu}

\begin{abstract}
XMM-Newton observed  SAX J2103.5+4545 on January 6, 2003, while
RXTE was monitoring the source. Using RXTE-PCA dataset
between December 3, 2002 and January 29, 2003, the
spin period and average spin-up rate
during the XMM-Newton observations were found to be
$354.7940\pm0.0008$ s and $(7.4\pm0.9)\times10^{-13}$Hz s$^{-1}$ respectively.
In the power spectrum of the 0.9-11 keV EPIC-PN lightcurve, we found 
quasi periodic oscillations around 0.044 Hz (22.7 s) with an rms fractional
amplitude $\sim $6.6 \%. We interpreted this QPO feature as the
Keplerian motion of inhomogenuities through the inner disk.
In the X-ray spectrum, in addition to the power law component with
high energy cutoff and $\sim6.4$ keV fluorescent iron emission line 
(Baykal et al., 2002), we discovered a soft component consistent with a 
blackbody emission with ${\rm{kT}}\sim1.9$keV. The pulse phase
spectroscopy of the source revealed that the blackbody flux peaked at the peak 
of the pulse 
with an emission radius $\sim 0.3$ km, suggesting the polar cap on the
neutron star surface as the source of blackbody emission. The flux of the iron 
emission line at $\sim 6.42$ keV was shown to peak at the off-pulse phase, 
supporting the idea that this feature arises from fluorescent emission of the 
circumstellar material around the neutron star rather than the hot region in 
the vicinity of the neutron star polar cap.
\end{abstract}

\section{Introduction}
The transient X-ray source SAX J2103.5+4545 was discovered by the 
Wide Field Camera on the {\it{BeppoSAX}} X-ray observatory during its outburst 
between 1997 February and September with 358.61s 
pulsations and a spectrum consistent with an absorbed power law 
model with the photon index of $\sim1.27$ and the absorption column 
density of $\sim3.1\times 10^{22}$cm$^{-2}$ 
(Hulleman, in't Zand, \& Heise 1998). 

After detection of another outburst in November 1999 by the 
{\it{all-sky monitor (ASM)}} on 
the {\it{Rossi X-ray Timing Explorer (RXTE)}},  the source was found to 
be active for more than a year, and was continously monitored through 
regular pointed {\it{RXTE}} observations. Using pulse arrival times, 
the orbital period and eccentricity of the orbit
were found to be 
12.68(25) days and 0.4(2)  (Baykal, Stark, \& Swank 2000a,b). 
In the timing analysis, the source was initially 
found to be spinning up for $\sim 150$ days, 
at which point the flux dropped quickly by 
a factor of $\simeq 7$, and a weak spin-down began afterwards 
(Baykal, Stark, \& Swank 2002). Strong  
correlation between X-ray flux and spin-up rate was explained by using
Ghosh \& Lamb (1979) accretion disk model. 
The X-ray spectra well fitted the
absorbed power law model with high energy cutoff and a $\sim $6.4 keV
fluorescent emission line (Baykal et al. 2002).
 
Orbital parameters found by using {\it{RXTE}} observations of the source 
(Baykal et al. 2000a,2000b) indicated that the source has a high mass
companion. Hullemann et al. (1998) pointed out a B8 type star within the 
BeppoSAX error box, but its distance ($\sim 0.7$ kpc) implied a luminosity 
too low to explain the spin-up that was seen in the RXTE observations. Recently, 
a possible candidate for the optical companion of SAX J2103.5+4545 with 
the visual magnitude of 14.2 was discovered 
(Reig \& Mavromatakis, 2003). 

SAX J2103.5+4545 was also observed with the {\it{INTEGRAL}} observatory in the
3-200 keV band with significant detection up to $\sim 100$ keV
(Lutovinov, Molkov,\& Revnivtsev 2003). The
spectral parameters found in the {\it{INTEGRAL}} observations of the source 
were found to be compatible with those found by Baykal et al. (2002).

Since the beginning of the most recent outburst in June 2002, SAX J2103.5+4545 
has been monitored continously by {\it{RXTE}} through regular pointed 
observations. It was possible to obtain some simultaneous 
coverage with the {\it{XMM-Newton}} observatory on January 6, 2003. 
The observation of {\it{XMM-Newton}} revealed 
a soft spectral component of the source which was well-represented by a 
blackbody model. This spectral model was verified by simultaneous fitting of 
January 6, 2003 {\it{RXTE}}-PCA observation. Using {\it{XMM-Newton}} dataset, 
we also discovered $\sim 22.7$s quasi periodic 
oscillations (QPO's) of this source for the first time. In this paper, we  
present our spectral and timing results of the analysis of RXTE and XMM 
datasets of SAX J2103.5+4545.  
 
\section{Observations}
\subsection{RXTE}
We analyzed RXTE observations of SAX J2103.5+4545 between December 3, 2002 
and January 29, 2003 with a total observation time of $\sim 52$ksec. 
This set of observations is a subset of the RXTE observations 
for the proposal number 70082.
The results presented here are based on data collected
with the Proportional Counter Array (PCA; Jahoda et al., 1996). The PCA 
instrument consists of an array 
of 5 proportional counters (PCU) operating in the 2-60 keV energy range, with 
a total effective area of approximately 6250 cm$^{2}$ and a field of view 
of $\sim 1^{\circ}$ FWHM. Although the number of active PCU's 
varied between 2 and 5 during the observations, our observations belong to the 
observational epoch for which background level
for one of the PCUs (PCU0) increased due to the fact that this PCU started
to operate without a propane layer. The latest combined background models
(CM) were used together with FTOOLS 5.2 to estimate the appropriate background. 

\subsection{XMM}
{\it{XMM-Newton}} observations took place on January 6, 2003 with an 
8.7 ksec continuous exposure.
Among the three EPIC detectors (Turner
et al. 2001; Str\"{u}der et al. 2001), the MOS1 and MOS2 detectors
were configured in Fast Uncompressed mode, while the PN was configured
in Fast Timing mode. Data collected by the EPIC detectors on
{\it{XMM-Newton}} were processed using version 5.4.1 of the
{\it{XMM-Newton}} Science Analysis System (SAS). 
We did not include the data collected by the
two Reflection Grating Spectrometers (RGS1 and RGS2) 
in data analysis since the count rates from these spectrometers were 
too low.  

\section{Data Analysis}

\subsection{Pulse Timing and Pulse profiles} 

In the timing analysis, we corrected the background subtracted
lightcurves of RXTE data to the barycenter of solar system.
The data have also been corrected for the orbit model using the eccentric 
orbital parameters given by Baykal et al. (2000b) with the new orbital epoch 
being MJD $52633.90\mp 0.05$. In order to estimate the pulse frequency and 
pulse frequency derivative accurately, we used $\sim 57.5$ days time span of 
RXTE observations between MJD 52611.48 and 
MJD 52668.90 which covers $\sim 8.7$ ksec short
observation of XMM data starting at MJD 52645.85.
We obtained the nominal pulse frequency 
by using Fourier transform and constructed 20 pulse profiles
(one pulse profile for each RXTE orbit) by folding the
lightcurve at this nominal pulse period. We found the pulse 
arrival times (phase offsets)
by cross-correlating these pulse profiles with a template chosen as the most
statistically significant pulse profile. In the pulse timing analysis, we used 
the harmonic representation
of the pulse profiles (Deeter \& Boynton 1985). In this technique, the pulse
profiles are expressed in terms of harmonic series and cross correlated
with the template pulse profile. The pulse phase offsets can be found of a
Taylor expansion, 
\begin{equation}
\delta \phi = \delta \phi _{0} + \delta \nu (t-t_{0}) +
\frac{1}{2} \dot \nu (t-t _{0})^{2} ,
\end{equation}
where $\delta \phi $ is the pulse phase offset deduced from the
pulse timing analysis, $t_{0}$ is the midtime of the observation,
$\phi_{0}$ is the phase offset at $t_{0}$, $\delta \nu$ is the
deviation from the mean pulse frequency (or additive correction to the
pulse frequency), and $\dot{\nu}$ is the pulsar's pulse frequency
derivative.
We fitted the phase offsets to the Taylor expansion. From
the fit, we found the pulse period corresponding to the 
XMM observation as $354.7940 \pm 0.0008$s and 57.5 days average spin-up rate 
as $(7.4 \pm 0.9) \times 10^{-13}$ Hz s$^{-1}$.
We did not see any significiant timing noise in the residuals of arrival times, 
which indicated that the spin up rate was stable through the observations.
The average 3-20 keV flux of RXTE observations was
$(5.5\mp 0.5) \times 10^{-10}{\rm{erg s}}^{-1}{\rm{ cm}}^{-2}$. The mean 
spin-up rate and the average X-ray flux were found to be consistent with 
the previously observed spin-up rate and
X-ray flux correlations during the 1999 outburst (see Fig. 7 in Baykal 
et al. 2002). Detailed timing noise of RXTE observations is in progress
and is not the scope of this paper (Baykal et al. 2004).

In Figure 1, we presented energy dependent pulse profiles of XMM-EPIC PN data. 
The feature peaking at the phase
$\sim 0.25$ before the peak of the main pulse was a prominent
feature of the pulse profile.  The pulse fraction was
$(50.9\mp 0.3)$\% at 0.9-11 keV, whereas it was found to be
slightly variable in the energy intervals shown in Figure 1 with a
minimum of $(48.5\mp 0.8)$\% at 7.5-11 keV and a maximum of
$(53.6\mp 0.7)$\% at 2.5-5.0 keV.

\subsection{Transient 22.7 sec QPO}

In the power spectra of the 0.9-11 keV EPIC-PN lightcurve we found 
quasi periodic oscillations around 0.044 Hz (see Fig. 2).
In order to test the significance of these oscillations,  
we averaged 9 power spectra and rebinned the frequencies by a factor 8.
Then we modeled the continuum power spectrum with a broken 
power law model with the break value $(4.45\mp 0.16)\times 10^{-2}$ Hz, and 
the power indices $-0.34\mp 0.08$ and $-2.14\mp 0.05$. 
We modeled the transient oscillations with a Lorentzian centered at
$(4.40\mp 0.12)\times 10^{-2}$ Hz, with the full width
half maximum (FWHM) of $(6.1\mp 2.0)\times 10^{-3}$ Hz.

To test the significance of this QPO feature, we normalized the 
power spectrum by dividing it by the continuum and we multiplied this result by 
2 (van der Klis 1989). The resultant power spectrum would be consistent
with a Poisson distribution for a degree of freedom $2\times 8\times 9$ = 144. 
As seen from the lower panel of Figure 2, there is a prominent peak at 
$\sim 0.044$ Hz with a minimum of excess power (including the error of power)
3.62 giving the total power as
$3.62\times 8\times 9$ = 260.64. This gives the probability of detecting 
a false signal ${\rm{Q}}(260.64|144)= 9.39\times 10^{-9}$.
Since we have 512 frequencies in each power spectra, total probability of 
having false signal becomes 
$9\times 512 \times 9.39\times 10^{-9} = 4.3\times 10^{-5}$. Therefore 
significance of transient oscillations is $1-4.3\times 10^{-5}\simeq0.99996$ 
which is consistent with more than 6 sigma level detection.   
The rms fractional amplitude associated with this QPO
feature was found to be $(6.6\mp 1.9)$\%. 

We searched transient oscillations using the RXTE-PCA lightcurves in the 3-20 
keV energy range,
however we did not see any significant transient oscillations.
Then we extracted a $\sim 50$\% portion of the overall XMM-EPIC PN lightcurve at 
the 3-10 keV energy range that  coincided
exactly with the $\sim 4.5$ ksec part of the RXTE PCA lightcurve on January 6,
2003 and performed power spectral analysis.  
We found that the significances of 0.044 Hz oscillations for RXTE and XMM 
lightcurves are $2.5\sigma$ and $2.8\sigma$ respectively.
We concluded that the QPO feature was originated mostly from the 
soft component of the spectrum or highly transient. Future observations 
are required to confirm these oscillations.

\subsection{Spectral Analysis}
We fitted overall background subtracted 1-10 keV 
spectra of PN, MOS1 and MOS2 to an absorbed power law model
(Morrison \& McCammon, 1983) with high energy cut-off (White, Swank, Holt 
1983). In addition, an iron line feature at 6.42 keV was required
in the spectral model (Case 1 in Table 1). However, this model
did not fit the spectrum of the source well, giving a reduced
$\chi^2$ of 2.69. Adding an additional blackbody component to the
model decreased the reduced $\chi^2$ to 1.23 (Case 2 in Table
1). Joint fit including the PCA data on January 6, 2003 and adding 2\%
systematic errors (see Wilms et al. 1999; Coburn et al. 2000), 
was possible to this model with a reduced
$\chi^2$ of 1.1 (Case 3 in Table 1 and Figure 3). Using Case 3 (i.e. the 
joint fit to the model including blackbody component), 1-20 keV unabsorbed 
flux was found to be $6.6\times 10^{-10}$ ergs s$^{-1}$ cm$^{-2}$.
Assuming a source distance of 3.2 kpc (Baykal et al. 2002), this value 
corresponds to a luminosity of $7.5\times 10^{35}$ ergs s$^{-1}$. To compare 
with the average X-ray flux, we found 3-20 keV flux of 
the January 6 observation to be $6.1\times 10^{-10}$ ergs s$^{-1}$ cm$^{-2}$.
This value is approximately 10\% greater than the 
57.5 days average 3-20 keV RXTE-PCA flux. This is reasonable since the   
XMM observation took place 3.55 days after the periastron passage where the 
X-ray flux reaches approximately its maximum (see Fig. 5 and 8 of 
Baykal et al. 2000b). It should be noted that line parameters 
obtained from 3-20 keV RXTE PCA data agree with those 
obtained from XMM EPIC data. However exclusion of the blackbody component, 
while fitting RXTE PCA data only, increases the absorption column density  
and equivalent width of line emission and make the power law index harder 
as shown in Table 1, case 4.

To study the spin-phase-resolved spectra using EPIC-PN
data, we divided the spin phase into 10 bins, and fitted 1-10 keV
spectrum of each bin with the model including blackbody component (i.e. 
model in case 3). Figure 4 is a plot of the spectral parameters as a function 
of the spin phase. For all the spin phases, we found that the model gives an 
iron line peak energy consistent with $6.42\pm0.04$ keV within 1$\sigma$, so 
we chose to fix this parameter. We also checked the consistence of freezing  
the cut-off energy, iron line sigma, and e-folding energy parameters, and 
found that these parameters did not vary significantly when they were 
thawed. From Figure 4, strong modulation 
(with a factor of $\sim 10$) of the blackbody flux with the spin phase is 
evident. Similarly, the flux of the power law
component was shown to be varying with the spin phase, but the 
variation was more moderate (with a factor of $\sim 3$) than the 
that of the blackbody component. 
The iron line feature at 6.42 keV was stronger for the off-pulse phases 
where the X-ray flux was lower.

\section{Discussion and Conclusion}
\subsection{QPO Feature of SAX J2103.5+4545}
Quasi-periodic oscillations in the X-ray band having periods in the 
range of
$\sim 2.5-100$s have been observed in many accretion powered X-ray pulsars:
4U 0115+63 (Soong \& Swank 1989), EXO 2030+375 (Angelini, Stella,\& Parmar 1989),
4U 1626-67 (Shinoda et al. 1990),
SMC X-1 (Angelini, Stella,\& White 1991), Cen X-3 (Takeshima et al. 1991),
V0332+53 (Takeshima et al. 1994), A0535+262
(Finger, Wilson,\& Harmon 1996), GRO J1744-28 (Zhang, Morgan,\& Jahoda 1996; 
Kommers et al. 1997),
X Per (Takeshima 1997), 4U 1907+09 (in't Zand, Baykal, \& Strohmayer 1998;
Mukerjee et al. 2001), XTE J1858+034 (Paul \& Rao 1998), 
LMC X-4, and Her X-1 (Moon \& Eikenberry 2001a,b).
The QPO feature that we found in the
XMM-Newton EPIC-PN light curve of SAX J2103.5+4545 which has a
peak period of $22.7\mp 0.6$s and fractional rms amplitude 
of $6.6 \pm 1.9$ percent is
quite typical (e.g. In't Zand et. al. 1998; Paul \& Rao 1998;
Takeshima et al. 1994).

Models that explain the QPO phenomenon in accretion
powered X-ray pulsars fall basically into three categories:
In the Keplerian frequency model, QPOs are produced due to some inhomogenuities 
at the inner edge of the Keplerian disk ($r_0$) and modulate the light curve 
at the Keplerian frequency $\nu _{QPO}=\nu_{K}$ 
(van der Klis et al. 1987). In the beat frequency model,  
the accretion flow onto the neutron star is modulated at the beat frequency 
between the Keplerian frequency at the inner edge of the accretion disk 
and the neutron star spin frequency 
$\nu _{QPO}=\nu_{K}-\nu _{s}$ (Alpar \& Shaham 
1985). The third model involves
accretion flow instabilities (Fronter, Lamb,\& Miller 1989; Lamb 1988),
and applies only to the sources that have luminosities close to Eddington 
limit, therefore it should not be applicable to our case for which the 
luminosity is well below the Eddington limit.

In our case,
QPO frequency $\nu _{QPO}= 4.4\times 10^{-2}$Hz is about one
order of magnitude greater than the spin frequency 
$\nu _{s}= 2.8185\times 10^{-3}$ Hz. 
Therefore, it is difficult  
to distinguish between a Keplerian model and a beat frequency model.

Assuming that the 22.7 s oscillation in SAX J2103.5+4545 is related to 
Keplerian orbital motion via either Keplerian 
frequency model or beat frequency model, and using the QPO and its FWHM values 
we obtain the radius of inner disk as 
\begin{equation}
r_{0}= \bigg(\frac{GM}{4\pi^2}\bigg)^{1/3}\nu_k^{-2/3} = 
(1.32^{+0.13}_{-0.11}) \times 10^{9} {\rm{cm}},
\end{equation} 
where M is 1.4 M$_{\odot}$ for a neutron star and G is the gravitational 
constant. 

From the strong correlation between 
pulse frequency derivatives and X-ray flux, Baykal et al. (2002) 
obtained for the distance to the source 
$3.2 \pm 0.8$ kpc and for the magnetic field $(12\pm 3)\times 10^{12}$ Gauss.
Using the distance and magnetic field values,
the inner edge of the Keplerian disk 
$r_{0}$ can be found as (Ghosh \& Lamb 1979)
\begin{equation}
r_{0}\simeq 0.52\mu^{4/7}(2GM)^{-1/7}\dot M^{-2/7} = 
(1.67_{-0.25}^{+0.23})\times 10^9 {\rm{cm}},
\end{equation}
where $\mu=BR^{3}$ is the neutron star magnetic moment with B the equatorial 
magnetic field, R the neutron star radius, and $\dot M$ the mass accretion rate 
having the value of $\simeq 4\times 10^{15}$g $s^{-1}$ for an accretion 
luminosity of $\simeq 7.5\times 10^{35}$ erg s$^{-1}$ (as estimated in Section 
3.3). 

The radius of the inner disk inferred from  
the Keplerian orbital motion of inhomogenuities 
and the one inferred from the Ghosh Lamb disk accretion model agree each other.
This shows that the idea that the QPOs are formed due to the Keplerian motion
of inhomogenuities is indeed promising as the 
explanation of the QPO of SAX J2103.5+4545 and the observed QPO frequency is 
consistent with the distance and the
magnetic field values estimated by Baykal et al. (2002).  
 
\subsection{Blackbody and Iron Line Features of the Energy Spectrum}

XMM-Newton observations of SAX J2103.5+4545 revealed for the first time that 
the energy spectrum of the source has a blackbody component peaking at 
$\sim 1.90$keV with the emission radius of $\sim 0.3$km.  
The blackbody radiation may come 
from the polar cap of the neutron star as it appears to for 
the Be/X-ray pulsar system EXO
2030+375 (Reig \& Coe 1999; Sun et al. 1994) and the millisecond X-ray pulsars 
SAX J1808.4-3658 (Gierlinski, Done,\& Didier 2002) and 
XTE J0920-314 (Juett et al. 2003). Blackbody emission radii on these X-ray 
pulsars are reported to be greater than $\sim 1$km.
The relatively high surface magnetic field of 
SAX J2103.5+4545 ($\sim 10^{13}$ Gauss) is
probably the reason for the relatively small blackbody emission radius ($\sim 
0.3$ km) compared to these X-ray pulsars. Although the contribution of 
blackbody component is relatively more
significant for lower energies (i.e. energies smaller than $\sim 3$ keV),
power law flux is $\sim 3$ times greater than the blackbody flux even 
at the 1-3 keV energy band.

In our case, it is unlikely that the blackbody emission comes from the 
reprocessed emission of the surrounding material or the accretion disk as in 
the case of Her X-1 (Endo et al. 2000), Cen X-3 (Burderi et al. 2000), 
SMC X-1 and LMC X-4 (Paul et al. 2002), 
since blackbody component in such cases is expected to be softer 
($kT\sim 0.1$keV). Lower blackbody temperature and smaller blackbody emission
radius at the off-pulse phase shown in Figure 4 are also indications of the
plausibility of the polar cap emission interpretation, as the regions of the 
soft polar cap emission must align with the peak of the
X-ray pulse of the pulsar.  

Using the spin-phase resolved spectroscopy, strength of the iron line feature 
at $\sim 6.42$keV was also found to vary significantly with the spin phase as 
seen in Figure 4. The peak energy of this feature clearly shows that it 
corresponds to the fluorescent iron K-line complex. This line complex feature 
is observed in the 
spectra of most of the X-ray pulsars (White et al. 1983; Nagase 1989) 
and is generally thought to be produced by the ions less ionized
than Fe XVIII in a relatively cool matter around the neutron star (e.g.
accretion disk, accretion disk corona) by fluorescent K$\alpha$ transition. 

Variation of the iron line feature with the spin phase can then be
interpreted as a sign that it is mainly produced outside
the polar cap region of the neutron star, thus should have a peak at
the off-pulse parts of the spin phase. From Figure 4, we see that 
iron line flux and iron line equivalent width vary strongly 
with the spin phase, peaking at the
off-pulse. Similar pulse phase dependence of the iron line feature is
also observed in Her X-1 (Choi et al. 1994).
  
\acknowledgments

We thank anonymous referee for helpful comments. We also thank 
Dr. T. Strohmayer and Dr. C. Markwardt for their useful 
discussions. S.\c{C}.\.{I}nam acknowledges the Integrated Doctorate Program 
scholarship from the Scientific and Technical Research Council of Turkey 
(T\"{U}B\.{I}TAK).

\clearpage
\begin{figure}
\epsscale{.80}
\plotone{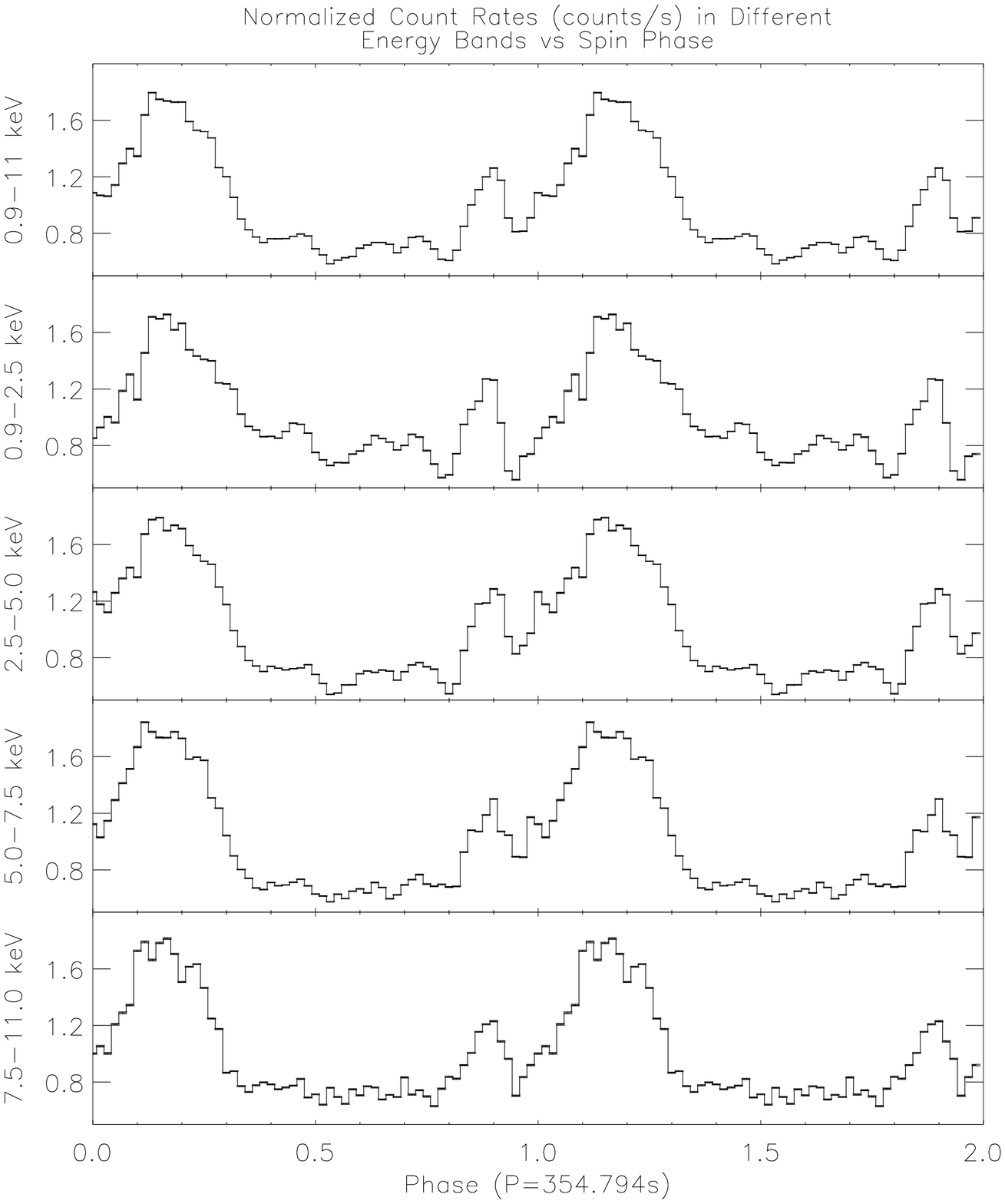}
\caption{Energy dependence of the pulse profiles of SAX J2103.5+4545. 
Pulse profiles were found by using EPIC-PN lightcurves that cover 
the energy ranges from top to bottom respectively.}
\end{figure}

\begin{figure}
\includegraphics[scale=.60]{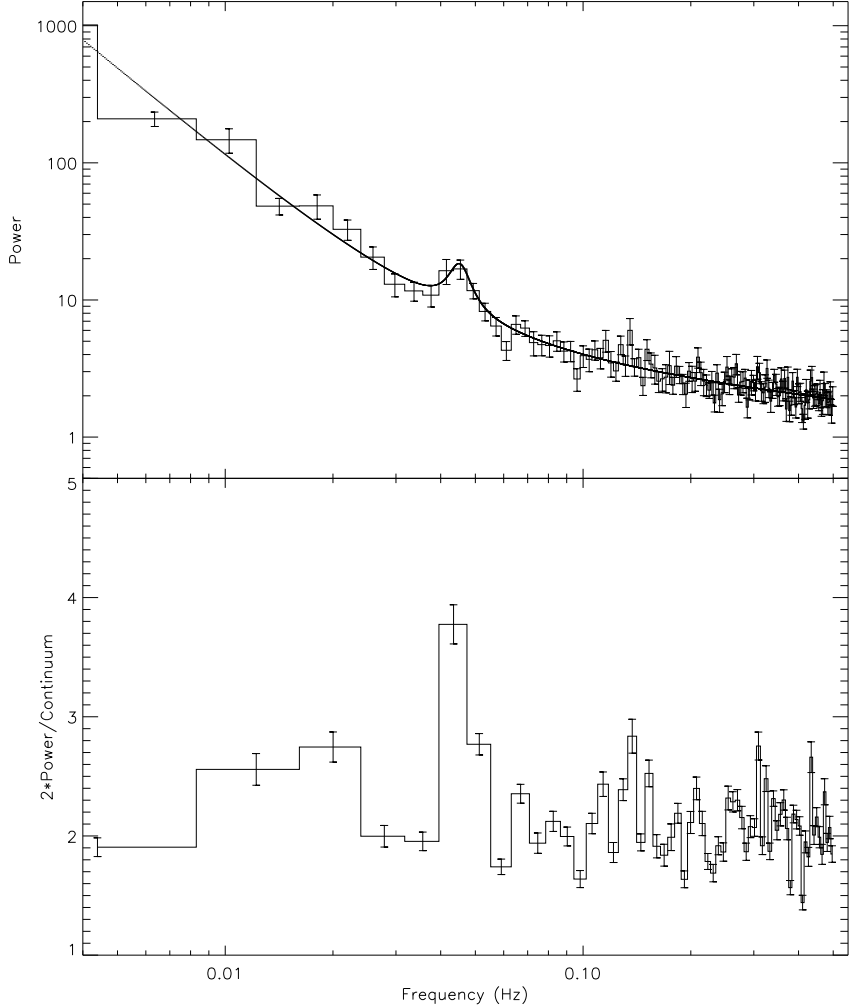}
\caption{{\bf{(top)}} Power spectrum obtained from the 0.9-11.0 keV EPIC-PN 
lightcurve and rebinned by a factor of 4. QPO feature centered 
at 0.044 Hz is the prominent feature of the power spectrum. 
{\bf{(bottom)}} 
Power spectrum rebinned by a factor of 8, multiplied by 2, and divided by the 
continuum fit consisting of a broken 
power law model with the power indices $-0.34\mp 0.08$ and $-2.14\mp 0.05$. 
Applying the method discussed by van der Klis (1989), significance of the QPO 
feature was calculated to be more than $6\sigma$ confidence level
using the value of the peak at $\sim 0.044$ Hz as shown in this plot.} 
\end{figure}
\clearpage
\begin{figure}
\begin{center}
\includegraphics[angle=-90,scale=.50]{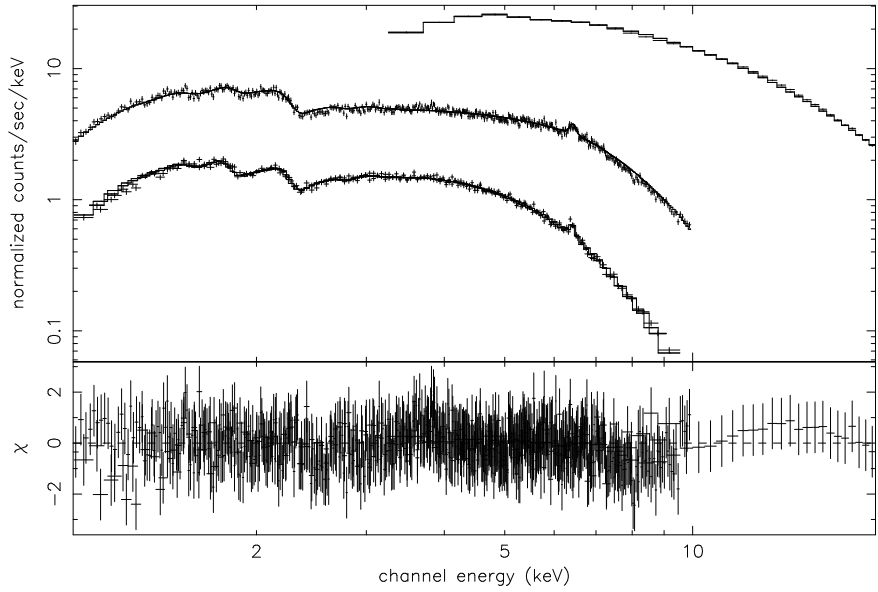}
\end{center}
\caption{1-20 keV combined PN, MOS1, MOS2 and PCA spectrum of SAX J2103.5+4545
observed on January 6, 2003. The bottom panel shows the residuals of the fit 
in terms of $\sigma$ values.}
\end{figure}
\clearpage
\begin{figure}
\epsscale{0.84}
\plotone{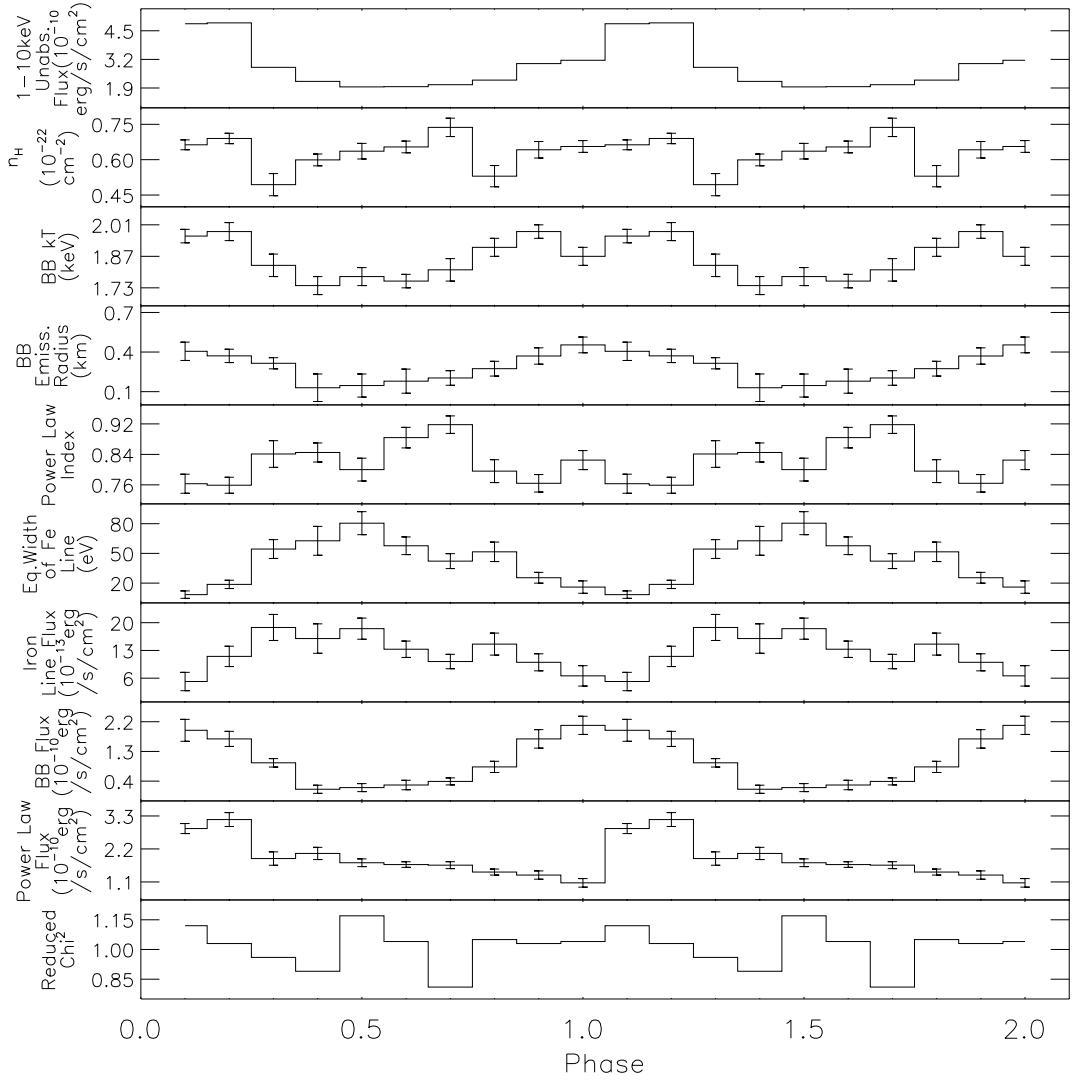}
\caption{Spin phase dependence of 1-10 keV X-ray flux in units of $10^{-10}$
ergs s$^{-1}$ cm$^{-2}$, Hydrogen column density in units 
of $10^{22}$ cm$^{-2}$, blackbody kT in 
units of keV, emission radius of the blackbody (assuming a 
source distance of 3.2 
kpc) in units of km, photon index of the power law component of the spectral
model, equivalent width of the iron line in units of eV, flux of the
iron line in units of $10^{-13}$ ergs s$^{-1}$ cm$^{-2}$, blackbody 
flux in units of $10^{-10}$ ergs s$^{-1}$ cm$^{-2}$, power law flux in units of 
$10^{-10}$ ergs s$^{-1}$ cm$^{-2}$, and reduced $\chi^2$. All of the errors show 
$1\sigma$ confidence level. The spectral model used in fitting consists
of an absorbed power law model with high energy cut-off, a soft blackbody 
component and an Iron line feature modeled as a Gaussian. 
The values of iron line peak energy, iron line 
sigma, cut-off and e-folding energies were found to be consistent
with the constant values 6.42 keV, 0, 
7.89 keV, and 27.1 keV respectively for all the spin phases, so these values
were kept fixed while the values of the other model parameters were 
found.}
\end{figure}
\clearpage
\begin{table}
\caption{Spectral Models of SAX J2103.5+4545}  
\tiny{
\begin{tabular}{l c c c c}\hline \hline
Parameter & Case 1 & Case 2 & Case 3 & Case 4 \\
& {\bf{(without BB)}} & {\bf{(with BB)}} & {\bf{(with BB)}} & {\bf{(without BB)}} \\
& (PN,MOS1,MOS2) & (PN,MOS1,MOS2) & (PN,MOS1,MOS2,PCA) & (PCA) \\
\hline
Multiplication Factors$^1$ & & &  \\
(1.00 for EPIC-PN)     & & & \\
------MOS 1 & $0.77\mp 0.01$ & $0.76\mp 0.01$ & $0.76\mp 0.01$ & na \\
------MOS 2 & $0.76\mp 0.01$ & $0.75\mp 0.01$ & $0.75\mp 0.01$ & na \\
--------PCA & na             & na & $1.26\mp 0.01$ & na \\
$n_H$ ($10^{22}$cm$^{-2}$) & $0.90\mp 0.02$ & $0.68\mp 0.01$ & $0.66\mp 0.02$ &
$2.98\mp 0.14$ \\
Iron Line Energy (keV) & $6.41\mp 0.04$ & $6.42\mp 0.02$ & $6.42\mp 0.02$ &
$6.36\mp 0.06$ \\
Iron Line Sigma (keV) & 0 (fixed) & 0 (fixed) & 0 (fixed) & $0.68\mp 0.13$ \\
Iron Line Equivalent Width & $48.0\mp 7.0$ & $37.1\mp 5.3$ & $36.5\mp 5.0$ &
$107\mp 15$ \\ 
(eV) & & & & \\
Iron Line Flux & $(1.69\mp 0.24)\times 10^{-12}$ & $(1.37\mp 0.20)\times
10^{-12}$ & $(1.36\mp 0.19)\times 10^{-12}$ & $(5.14\mp 0.72)\times 10^{-12}$ \\
(ergs s$^{-1}$ cm$^{-2}$) & & & \\ 
Iron Line Normalization & $(1.65\mp 0.24)\times 10^{-4}$ & $(1.33\mp 0.19)\times 
10^{-4}$ 
& $(1.32\mp 0.18)\times 10^{-4}$ & $(5.00\mp 0.70)\times 10^{-4}$ \\
(photons cm$^{-2}$ s$^{-1}$) & & & \\
Blackbody kT (keV) & na & $1.91\mp 0.04$ & $1.88\mp 0.02$ & na \\
Blackbody Normalization & na & $0.88\mp 0.04$ & $0.83\mp 0.07$ & na \\
(km$^2$ (10kpc)$^{-2}$) & & & \\
Power Law Index & $0.83\mp 0.02$ & $0.82\mp 0.04$ & $0.77\mp 0.05$ & $1.14\mp
0.05$ \\
Power Law Normalization & $(1.61\mp 0.32)\times 10^{-2}$ & $(1.03\mp 0.26)\times 10^{-2}$
& $(1.01\mp 0.27)\times 10^{-2}$ & $(3.88\mp 0.07)\times 10^{-2}$ \\
(photons keV$^{-1}$ cm$^{-2}$ s$^{-1}$) & & & \\
Cut-off Energy (keV) & 7.89 (fixed) & 7.89 (fixed) & 7.89 (fixed) & 7.89 (fixed)
\\
E-folding Energy (keV) & 27.1 (fixed) & 27.1 (fixed) & 27.1 (fixed) & 27.1
(fixed) \\
Reduced $\chi^2$ & 2.69 (1259 d.o.f) & 1.23 (1253 d.o.f) & 1.11 (1283 d.o.f) &
1.20 (30 d.o.f) \\
\hline
\end{tabular}}\\
\tiny{{\bf{$^1$ For the cases 1, 2 and 3, we multiplied the 
entire model with a factor
which is varying with the instrument to account for the different 
normalizations of the instruments (the value of the constant was fixed
to be 1 for EPIC-PN).}}}
\end{table}

\end{document}